\documentclass[twocolumn,showpacs,prl]{revtex4}
\usepackage{amssymb}
\usepackage{graphicx}
\usepackage{dcolumn}
\usepackage{bm}

\begin{document}

\title{On radiative damping in plasma-based accelerators}
\author{I.~Yu.~Kostyukov}
\email{kost@appl.sci-nnov.ru}
\author{E.~N.~Nerush}
\author{A.~G.~Litvak}
\affiliation{Institute of Applied Physics, Russian Academy of Sciences, 603950 Nizhny Novgorod, Russia}


\begin{abstract}
Radiative damping in plasma-based electron accelerators is analyzed.
The electron dynamics under combined influence of the
constant accelerating force and the classical radiation reaction force
is studied. It is shown that electron acceleration cannot be limited
by radiation reaction. If initially the accelerating force was
stronger than the radiation reaction force then the electron acceleration
is unlimited. Otherwise the electron is decelerated by radiative damping
up to a certain instant of time and then accelerated without limits.
Regardless of the initial conditions the infinite-time asymptotic
behavior of an electron is governed by self-similar solution providing
unlimited acceleration. The relative energy spread induced by the radiative
damping decreases with time in the infinite-time limit. 
\end{abstract}

\maketitle

The plasma-based methods of electron acceleration demonstrate an impressive
progress in the last ten years. The quasimonoenergetic electron bunches
are generated in laser-plasma acceleration experiments \cite{Mangles2004}.
The electron energy in laser wakefield acceleration experiments exceeds
$1$~GeV for cm-scale acceleration length \cite{Leemans2006} and
energy doubling of $42$~GeV electrons in a meter-scale plasma wakefield
accelerator is demonstrated \cite{Blumenfeld2007}. Recently the physics
of linear colliders based on laser-plasma accelerators have been discussed
\cite{Schroeder2010,Nakajima2011}.

The accelerating structure in the plasma-based methods is a plasma
wave generated behind the driver which can be the laser pulse or the
electron bunch. There is a number of effects which limit the energy 
gain in the plasma-based accelerators \cite{Esarey2009}. One of the main
limitations comes from the dephasing. The velocity of the relativistic
electrons becomes slightly higher than the plasma wave phase velocity,
which is determined by the driver velocity. The accelerated
electrons slowly outrun the plasma wave and leave the accelerating
phase. This problem can be partially solved by the use of proper longitudinal
gradient of plasma density \cite{Katsouleas1986,Pukhov2008}. Another
limitation is caused by the driver depletion as the driver energy
converts into the energy of the plasma wave. The driver evolution
during acceleration (e.~g. laser pulse diffraction or electron bunch 
expansion) also imposes certain restrictions on
the electron energy gain. In the case of laser-plasma accelerators
the laser pulse can be guided over long distances in the preformed
plasma density channel \cite{Esarey1994} or with relativistic optical
guiding when diffraction is compensated by relativistic self-focusing
\cite{Litvak1969}. In general, in order to accelerate electrons far
beyond the energy limited by these effects the multistage schemes
can be used. 

The electron acceleration in the plasma wave is accompanied with the
transverse betatron oscillations caused by the action of the focusing
force on the electron from the plasma wakefield. The accelerating
force and the focusing force acting on the relativistic electron near
the driver axis can be approximated as follows $F_{acc}=fmc\omega_{p}$
and $F_{\bot}\simeq-m\kappa^{2}\omega_{p}^{2}r$, respectively, where
$r$ is the transverse displacement of the electron from the driver
axis, $f$ and $\kappa$ are the numerical factor and the focusing
constant, respectively, determined by the parameters of the driver
and the plasma, $\omega_{p}=\left(4\pi e^{2}n/m\right)^{1/2}$ is
the plasma frequency, $n$ is the density of the background plasma,
$m$ and $e=-\left|e\right|$ are the electron mass and the electron
charge, respectively \cite{Esarey2009}. For example, if the driver
is the linearly polarized Gaussian laser pulse with resonant pulse
duration then $f=0.35a_{0}^{2}\simeq0.7$ and $\kappa^{2}\simeq0.11$, 
where $a_{0}=eE_{L}/(mc\omega_{L})=2^{1/2}$
is chosen, $E_{L}$ is the laser field amplitude, $\omega_{L}$ is
the laser frequency \cite{Nakajima2011}. The period of the betatron oscillations is 
$\omega_{\beta}=\omega_{p}\kappa\gamma^{-1/2}$,
where $\gamma$ is the relativistic gamma-factor of the electron.

The electrons undergoing betatron oscillations emit synchrotron radiation
\cite{Esarey2002,Kostyukov2003}. The radiated power can be estimated
as follows $P_{rad}\simeq2r_{e}\gamma^{2}F_{\bot}^{2}/(3mc)$, where
$r_{e}=e^{2}/(mc^{2})\simeq3\cdot10^{-13}$~cm is the classical electron
radius, $c$ is the speed of light. Since the power is proportional
to the square of the electron energy, the radiation losses can stop
electron acceleration at some threshold value of the electron energy.
The threshold energy can be estimated by balancing the accelerating
force and the radiation reaction force, $F_{rrf}\simeq P_{rad}/c$,
so that $\gamma_{th}^{2}\simeq f/(\epsilon\kappa^{4}R_{\beta}^{2})$,
where $R_{\beta}=k_{p}r$ is the normalized amplitude of betatron
oscillations, $\epsilon=2r_{e}\omega_{p}/(3c)$ and $k_{p}=\omega_{p}/c$.
The threshold energy is $\sim 100 \:$GeV for $f=0.7$, $n=10^{19}\,\mbox{cm}^{-3}$
and $R_{\beta}=1$ and $\kappa^{2}=0.11$. Therefore the radiative damping may be a serious
limitation of electron acceleration. 

The electron acceleration in plasma with the radiation reaction effect
has been studied theoretically \cite{Schroeder2010,Nakajima2011,Michel2006,Kostyukov2006}.
The radiation reaction has been treated as a perturbation \cite{Michel2006}.
The first-order radiative correction to the energy gain of the accelerated
electron bunch and the energy spread induced by radiation emission
have been derived for the constant accelerating force. The dependence
of the electron energy on time has been calculated in the plasma channel
without the accelerating force and with the radiation reaction force \cite{Kostyukov2006}.
Here we study the electron acceleration treating the radiation damping
unperturbatively and analyzing the infinite-time limit.

We start from the relativistic equation for electron motion in an
electromagnetic field with the radiative reaction force in Landau-Lifshits
form \cite{Landau2} \begin{eqnarray}
\gamma\frac{du^{i}}{dt}=\frac{c r_{e}}{e}F^{ik}u_{k}+\frac{2r_{e}^{2}}{3mc}F_{rad}^{i},
\label{reaction}
\end{eqnarray}
 where $F_{rad}^{i}=F_{1}^{i}+F_{2}^{i}+F_{3}^{i}$, $F_{1}^{i}=
 \left(e/r_{e}\right)\left(\partial F^{ik}/\partial x^{l}\right)u_{k}u^{l}$,
$F_{2}^{i}=-F^{il}F_{kl}u^{k}$, $F_{3}^{i}=
\left(F_{kl}u^{l}\right)\left(F^{km}u_{m}\right)u^{i}$,
$F_{ik}$ is the electromagnetic field tensor, $u_{k}$ is the 4-velocity
of the electron. The first term in Eq.~(\ref{reaction}) corresponds
to the Lorentz force and the last term corresponds to the radiation
reaction force. We assume that the ultrarelativistic electrons ($\gamma\gg1$)
are accelerated along $x$-axis by the force 
$F_{acc}\gg F_{\bot} v_{\bot}/c$
and undergo betatron oscillations driven by the focusing force 
$F_{\bot}\simeq-m\kappa^{2}\omega_{p}^{2}y$.
Under our assumptions, $F_{3}\gg F_{1},\: F_{2}$ and the focusing
forces make a major contribution to the energy losses through radiation.
It is convenient to introduce new variables 
$P=(p_{y}/mc)\epsilon^{1/2}f^{1/2}$,
$Y=yk_{p}f^{3/2}\epsilon^{1/2}$, $T=\omega_{p}t\kappa^{2}/f$, $G=\gamma\kappa^{2}f^{-2}$.
Then Eq.~(\ref{reaction}) can be reduced to the form

\noindent \begin{eqnarray}
\frac{dP}{dT}=-Y-Y^{2}PG,\label{approx11}\\
\frac{dY}{dT}=\frac{P}{G},\label{approx21}\\
\frac{dG}{dT}=1-Y^{2}G^{2},\label{approx31}\end{eqnarray}
 The obtained equations describe the betatron oscillations with the radiative
damping. The first term on the right-hand side of Eq.~(\ref{approx31})
describes the action of the accelerating force, while the second term
describes the radiative damping.

When the number of the betatron oscillations is large, we can use the
averaging method~\cite{Bogolubov}. To do this let us introduce a
new variable, $S$, so that $2S=\left|U\right|^{2}=Y^{2}+P^{2}/G
=R_{\beta}^{2}f^{3}\epsilon\simeq2\left\langle Y^{2}\right\rangle $
and $U\exp\left(i\int G^{-1/2}dT\right)=Y-iG^{-1/2}P.$ After averaging
over the fast time related to the betatron oscillations the averaged
equations are \begin{eqnarray}
\frac{dS}{dT}=-\frac{1}{2}\frac{S}{G}-\frac{1}{4}GS^{2},\label{bs1}\\
\frac{dG}{dT}=1-SG^{2}.\label{bs2}\end{eqnarray}
\noindent
As $G>0$ and $S>0$ then $dS/dt<0$ and the amplitude of the betatron
oscillations always decreases with time. This means that for arbitrary
electron energy the betatron oscillation amplitude will be small enough
at certain instance of time to be radiation reaction force 
less than the accelerating force.

At the absence of the accelerating force ($f=0$), it follows from
Eqs.~(\ref{bs1}) and (\ref{bs2}) that $SG^{-1/4}=\textrm{const}$
and $\gamma=\gamma_{0}\left(1+5\epsilon R_{\beta , 0}^{2}\gamma_{0}\omega_{p}t/16\right)^{-4/5}$,
which is in agreement with the solution calculated in Ref.~\cite{Kostyukov2006}, 
where $R_{\beta , 0} = R_{\beta } (t=0) $.
At the absence of the radiation reaction (the last terms in RHS of Eqs.~(\ref{bs1})
and (\ref{bs2}) are absent) we get $G=G_{0}+T$, $\sqrt{G}S=\textrm{const}$.
The radiation reaction effect can be treated as a perturbation. To the first order in the radiation
reaction force the normalized electron energy is 
$G=G_{0}+T-(2/5)\left[1-\left(G_{0}+T\right)^{5/2}\right]$,
which is in agreement with the result obtained in Ref.~\cite{Michel2006}.

\begin{figure}
\includegraphics[width=8cm]{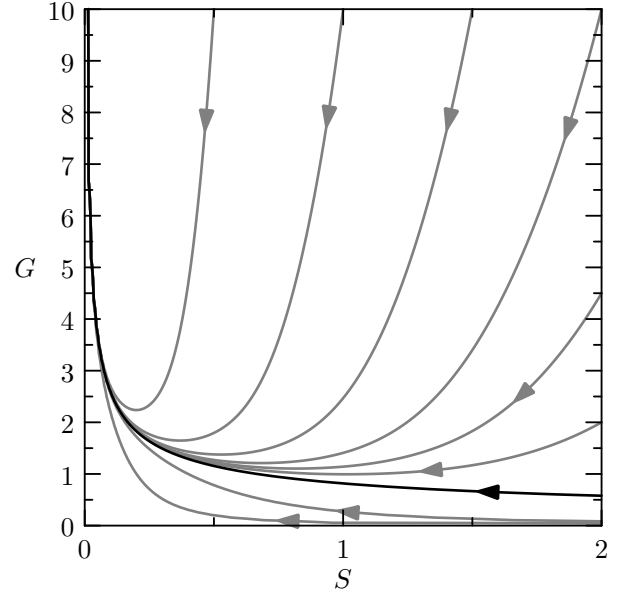} \caption{The phase portrait of 
the system governed by Eqs.~(\ref{bs1}) and
(\ref{bs2}).}
\label{fig1} 
\end{figure}

The system of Eqs.~(\ref{bs1}) and (\ref{bs2}) has integral
of motion \begin{equation}
I=\frac{1-3 S G^2 /2}{S^{9/4} \left(S G^2 \right) ^ {3/4}} =\textrm{const}.
\label{integral1}
\end{equation}
\noindent The electron trajectories in the phase space $S-G$ are the
integral lines determined by Eq.~(\ref{integral1}). The phase portrait
of the system governed by Eqs.~(\ref{bs1}) and (\ref{bs2}) is shown
in Fig.~\ref{fig1}. It is seen from Fig.~\ref{fig1} that if initially
the accelerating force is stronger than the radiation reaction force
($SG^{2}<1$) then the electron energy monotonically increases with
time. Otherwise the electron energy decays up to the time instance
when $F_{acc}=F_{rrf}$ (that corresponds to $SG^{2}=1$) and then
it monotonically increases with time. It is also seen from Fig.~\ref{fig1}
that all electron trajectories merge in the the limit $t\rightarrow\infty$
so that $G\rightarrow\infty$ and $S\rightarrow0$. 
It follows from Eq.~(\ref{integral1}) that $S=2G^{-2}/3$ in this limit.
We will call the electron acceleration in this 
limit as an asymptotic acceleration regime (AAR). 

\noindent %

\begin{figure}
\includegraphics[width=8cm]{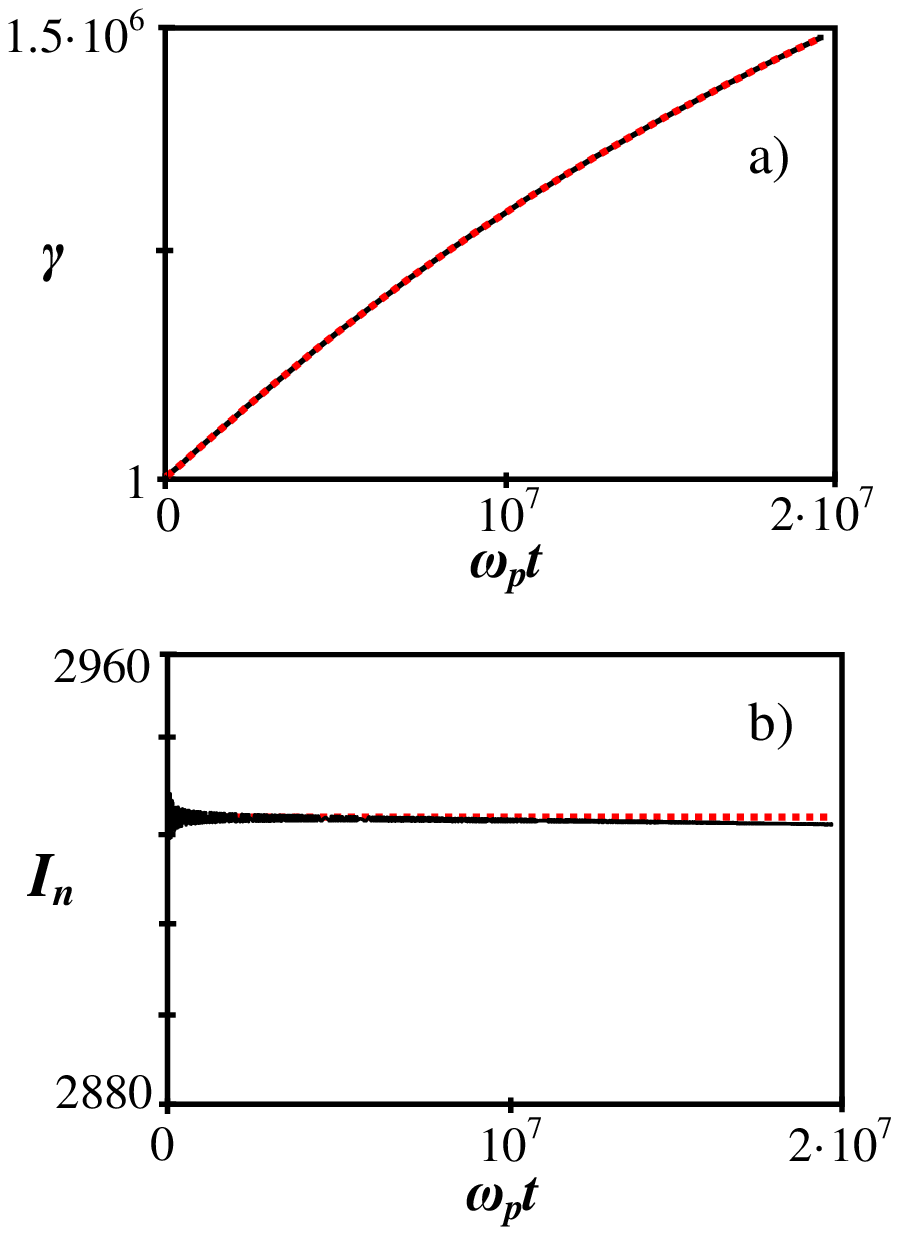} 
\caption{The dependence of a) $\gamma$  and b) $I_{n}$ on $\omega_{p}t$
calculated by solving of the exact Eq.~(\ref{reaction}) (black solid
lines) and by solving of the approximate Eqs.~(\ref{bs1})-(\ref{bs2})
(red dashed lines) for $f=0.1$, $\kappa^{2}=0.5$, $n=10^{15}$ $\,\mbox{cm}^{-3}$ 
and for initial conditions $\gamma_{0}=2000$, $R_{\beta , 0}=0.8$, $p_{y,0}=0$.}
\label{fig2} 
\end{figure}

We verify our analytical results by numerical simulations.
The exact equation (\ref{reaction}) and the averaged equations of
motions~(\ref{bs1}) and (\ref{bs2}) are integrated numerically
for test electrons for $f=0.1$ and $n=10^{15}\,\mbox{cm}^{-3}$.
For simplicity, we consider the structure of the transverse electromagnetic
field similar to the bubble regime: $\kappa^{2}=0.5$ and 
$E_{\bot}\approx H_{\bot}$. The dependence of the
normalized integral of motion $I_{n}=I^{-1}(\epsilon\kappa f^{2})^{-3}$,
and $\gamma$ on $\omega_{p}t$ for initial condition $\gamma_{0}=2000$
and $R_{\beta ,0}=0.8$, $p_{y,0}=0$ is shown on Fig.~\ref{fig2}.
It is seen from Fig.~\ref{fig2} that the solution of the exact
equations and that of the approximate averaged are in a good agreement.
Moreover, the integral $I$ is almost constant for the exact equations
(\ref{reaction}) (see Fig.~\ref{fig2}c).

\noindent %

We can introduce new variables $g=G/G_{tr}$, $\tau=T/T_{tr}$ and
$s=\left(S/S_{tr}\right)\left(G/G_{tr}\right)^{-1/4}$, where $G_{tr}=
T_{tr}=S_{tr}^{-2} = I^{2/9}$.
Then Eqs.~(\ref{bs1}), (\ref{bs2}) and (\ref{integral1}) are
reduced to the form which does not depend on any parameters. Therefore
the characteristic time of transition to AAR is $\sim T_{tr}$.
The solution of the equations can be written in term of hypergeometric function,
$\,_{2}F_{1}\left(a,b;c;z\right)$, ~\cite{Abramowitz} as follows $\varphi\left(s\right)
-\varphi\left(s_{0}\right)= \tau$,
$\varphi\left(s\right)=2^{4/9}\left(3+2s^{2}\right)^{5/9}s^{-4/9}-2^{13/9} 3^{5/9} 
s^{14/9}\,_{2}F_{1} \left(7/9,4/9;16/9;-2s^{2} /3\right)$, where 
$s_{0}=s\left(\tau=0\right)$. The asymptotic expansions of function $\varphi(s)$ are 
$\varphi\left(s\right)\approx 3 (3s/2)^{-4/9}$ for $s\ll 1$, 
$\varphi\left(s\right) \approx \delta +  s^{-4/3}$
for $s\gg 1$, where $\delta \approx1.85$.
Thus in the limit $\tau\gg1$ $s\sim\tau^{-9/4}\ll1$ and $g\sim\tau\gg1$.

To derive the asymptotic solution the initial condition
should be applied. We assume that $S_{0}G_{0}^{2}\ll1$ (so that $s_0 \ll 1$
 and $I \simeq S^{-3}G^{-3/2}$)
which is typical for the initial parameters of the electron beam.
For example, this condition is fulfilled for the initial parameters 
$\gamma_{0}mc^{2} < 0.1$~TeV, $n<10^{18}$~cm$^{-3}$, $R_{\beta , 0}=1$,
$f=0.7$, $\kappa^{2}=0.11$. Making of use the asymptotic expansion for
$s\ll1$ and $s_{0}\gg1$ we have $(9/4)s^{-4/9}\approx\tau+\delta$.
Therefore the normalized electron energy and the square of the normalized
betatron amplitude are in the limit $T\gg T_{tr}$ 
\begin{eqnarray}
G = \frac{\delta}{3}G_{tr}+\frac{1}{3}T, &  & 
S = \frac{2}{3}G^{-2}.
\label{gas}\end{eqnarray}
\emph{We can conclude that in AAR $F_{rrf}=2F_{acc}/3$ 
so that the electron energy increases linearly with time while the
betatron amplitude is reversely proportional to the time.}

\begin{figure}
\includegraphics[width=8cm]{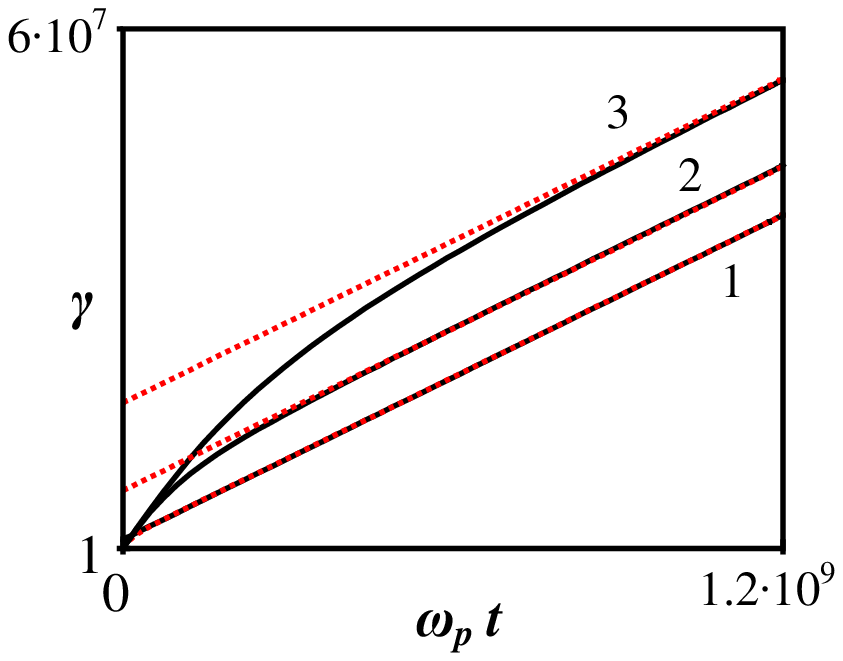} \caption{The dependence of $\gamma$ on 
$\omega_{p}t$ in AAR: analytic solution
(red dashed lines) and numerical solution (black solid line) for $R_{\beta , 0}=0.8$
(lines 1), $R_{\beta , 0}=0.2$ (lines 2) and $R_{\beta , 0}=0.1$ (lines 3). The
other parameters are the same as in Fig.~\ref{fig2}.}
\label{fig3} 
\end{figure}

The averaged equations of motions~(\ref{bs1}) and (\ref{bs2}) are
integrated numerically for the test electrons with the same parameters
as for Fig.~\ref{fig2} for three values of the initial betatron
amplitude $R_{\beta , 0}=0.8,\:0.2,\:0.1$. It is seen from Fig.~\ref{fig3}
that the asymptotic solution (\ref{gas}) is in a good agreement with
the result of numerical integration.

The radiation damping rate varies for the electrons with different betatron
oscillation amplitudes. This causes the energy spread in the electron
bunch accelerated in the plasma wave. We assume that the amplitude
of the betatron oscillations of the electrons in the accelerated bunch
is uniformly distributed in the range $R_{min}<R_{\beta , 0}<R_{max}$ and
$R_{max}\gg R_{min}$. We also again assume that $S_{0}G_{0}^{2}\ll1$. 
Then the normalized mean energy and the normalized square
of the relative energy spread are in AAR 
\begin{eqnarray}
\left\langle G\right\rangle \simeq \frac{2}{R_{max}^{2}}
\intop_{R_{min}}^{R_{max}}G R_{\beta , 0} dR_{\beta , 0} 
\simeq G_{max} 
\delta +\frac{T}{3},
\label{g}\\
\sigma_{G}^{2} = \left\langle G^{2}\right\rangle -\left\langle G\right\rangle ^{2}
\simeq G_{max}^2 \frac{\delta ^2}{3} \left( \frac{R_{max} }{R_{min}}
\right)^{2/3},
\label{dg}
\end{eqnarray}
\noindent where $G_{max}=G_{tr}(R_{\beta , 0}=R_{max})$. It follows from 
Eqs.~(\ref{g}) and (\ref{dg}) that the relative energy 
spread, $\sigma_{G} /\left\langle G\right\rangle $, decreases 
with time in AAR.

\noindent %

Eqs.~(\ref{approx11})-(\ref{approx31}) are derived under conditions 
that $F_{\bot}$ gives the main
contribution to the radiative damping and $F_{3}\gg F_{1},\: F_{2}$.
However $F_{\bot}$ goes to zero in the limit $t\rightarrow\infty$. Therefore we
should check: should the accelerating force and terms $F_{1}$, $F_{2}$
 be taken into account in the radiation reaction force in this limit?
First it is significant that the radiation reaction force remains
constant in AAR because $F_{\bot} \sim R_{\beta}\rightarrow0$ and $\gamma\rightarrow\infty$
for $t\rightarrow\infty$ in such way that $R_{\beta}^{2}\gamma^{2}=\mathrm{const}$.
Making of use Eq.~(\ref{gas}) and relation $v_{y}\sim\omega_{\beta}y$
we get $F_{2}/F_{3}\sim f\epsilon\ll1$ and 
$F_{1}/F_{3}\sim(3/4)\kappa^{2}f\gamma^{-1/2}\epsilon^{1/2}\ll1$,
where we assume that $\kappa\sim f\sim1$. The contribution from the
accelerating force (or from $E_x$) to $F_{3}$ is of the order
$F_{2}/F_{3}\ll1$. Therefore our model defined by Eqs.~(\ref{bs1})
and (\ref{bs2}) is valid in AAR. For high energy electrons
 quantum electrodynamics (QED) effects can be
important. The energy of the photon emitted by the accelerated electron
can be so high that the quantum recoil becomes strong. The photon emission
can be treated in classical approach if QED parameter 
$\chi=\left[\left(mc\gamma\mathbf{E}+\mathbf{p}\times\mathbf{H}\right)^{2}-
\left(\mathbf{p}\cdot\mathbf{E}\right)^{2}\right]^{1/2}/\left(mcE_{cr}\right)
\simeq\gamma F_{\perp}/(eE_{cr})$
is much less than unity, where $E_{cr}=m^{2}c^{3}/(e\hbar)\approx1.32
\times10^{16}\:$V/cm
is the QED critical field \cite{Landau4}. $\chi$ can be estimated
in AAR as follows $\chi\approx\left[\left(2f/\alpha\right)
\left(\hbar\omega_{p}/mc^{2}\right)\right]^{1/2}\ll1$,
where $\alpha=e^{2}/\hbar c\approx1/137$ is the fine structure constant.
Therefore the classical approach for the radiation reaction force is valid
in the limit $t\rightarrow\infty$ because, like for the corrections
to the radiation reaction force, the growth of $\gamma$ in $\chi$
is compensated by decreasing of $F_{\bot}$.

The distance passed by the electron before reaching AAR is $ k_p l_{tr} \simeq
 \left(f/\kappa^{2}\right)T_{tr}\simeq 1.6
\left( \epsilon^{2} \gamma_{0} 
 R_{\beta , 0}^{4}f \kappa^{8} \right)^{-1/3}$.
For the initial parameters $n=10^{18}$~cm$^{-3}$, $R_{\beta,0}=1$,
$\gamma_{0}=2\cdot10^{3}$, $f=0.7$, $\kappa^{2}=0.11$ the electron
comes into AAR after passing $7800$ laser-driven acceleration stages
with total distance $l_{tr}\simeq73$~m, achieving the energy 
$\gamma mc^{2}\simeq 5$~TeV
and $R_{\beta}\simeq0.008$, where the stage distance is
chosen to be equal to the half dephasing length \cite{Nakajima2011}
and the distance between the acceleration stages is neglected. For
the rarefied plasma $n=10^{15}$~cm$^{-3}$, AAR is achieved in $78$
stages with $l_{tr}\simeq 23$~km, $\gamma mc^{2}\simeq48$~TeV and
$R_{\beta}\simeq 0.005$. AAR may be achieved within one
acceleration stage in the proton-driven acceleration schemes because
of very large dephasing length \cite{Cadwell2009}. 

In conclusions, we have shown that the electron acceleration is not limited
by the radiative damping in plasma-based accelerators. Even if the
radiation reaction force is stronger than the accelerating force at
the beginning, then acceleration eventually succeeds deceleration with time.
The damping of the betatron oscillations leads to the transition
to the self-similar asymptotic acceleration regime in the 
infinite-time limit when the radiation reaction force becomes 
equal to two thirds of the accelerating force. The relative 
energy spread induced by the radiative damping in the accelerated 
electron bunch decreases with time in this regime. 
This opens possibility to use high density plasma at the late stages
of multistage plasma-based accelerators despite the fact that the
radiative damping is enhanced as density increases. The high density
plasma can be favorable because it provides high accelerating gradient
and, thus, reduces the length of the acceleration stages. 
The obtained results can be also applied
to any other accelerating systems with the linear focusing forces. 

This work was supported in parts by the Russian Foundation for Basic
Research, the Ministry of Science and Education of the Russian Federation,
the Russian Federal Program ``Scientific and scientific-pedagogical
personnel of innovative Russia''.

\end{document}